\documentclass[twocolumn,showpacs,preprintnumbers,amsmath,amssymb,nofootinbib]{revtex4-1}

\usepackage{graphicx}

\usepackage{bm}

\usepackage{amssymb}
\usepackage{amsfonts}
\usepackage{amsmath}

\usepackage{color}
\usepackage[mathscr]{eucal}

\definecolor{nv}{rgb}{0.1,0.1,0.6}
\definecolor{pr}{rgb}{0.2,0.1,0.5}
\definecolor{mg}{rgb}{0.4,0.0,0.4}

\newcommand{\nn}{\nonumber}

\newcommand{\beq}{\begin{equation}}
\newcommand{\eeq}{\end{equation}}
\newcommand{\beqy}{\begin{eqnarray}}
\newcommand{\eeqy}{\end{eqnarray}}
\newcommand{\beqyn}{\begin{eqnarray*}}
\newcommand{\eeqyn}{\end{eqnarray*}}
\newcommand{\nl}{\newline}

\newcommand{\bs}{\begin{slide}}
\newcommand{\es}{\end{slide}}
\newcommand{\bc}{\begin{center}}
\newcommand{\ec}{\end{center}}
\newcommand{\bmin}{\begin{minipage}}
\newcommand{\emin}{\end{minipage}}

\newcommand{\mc}{\mathcal}

\newcommand{\bi}{\begin{itemize}}
\newcommand{\ei}{\end{itemize}}

\newcommand{\ra}{\rangle}

\usepackage{latexsym}

\usepackage{dsfont}
\usepackage{multirow}

\newcommand{\bea}{\begin{eqnarray}}
\newcommand{\eea}{\end{eqnarray}}
\newcommand{\be}{\begin{equation}}
\newcommand{\ee}{\end{equation}}

\newcommand{\ud}{\mathrm{d}}

\newlength\savedwidth

\newcommand{\uvec}[1]{\boldsymbol{#1}}

\newcommand{\pure}{\text{pure}}
\newcommand{\phys}{\text{phys}}

\begin{document}

\title{Resolution of a conflict between Laser and Elementary Particle Physics}

\author{Elliot Leader}
 \email{e.leader@imperial.ac.uk}
\affiliation{Blackett laboratory \\Imperial College London \\ Prince Consort Road\\ London SW7 2AZ, UK}

\date{\today}
\begin{abstract}
The claim some years ago, contrary to all textbooks, that the angular momentum of a photon (and gluon) can be split in a gauge-invariant way into an orbital and spin term, sparked a major controversy in the Particle Physics community. A further cause of upset was the realization that the  gluon polarization in a nucleon, a supposedly physically meaningful quantity, corresponds only to the gauge-variant gluon spin derived from  Noether's theorem, evaluated in a particular gauge. On the contrary, Laser Physicists have, for decades,  been happily measuring physical quantities which correspond to orbital and spin angular momentum evaluated in a particular gauge. This paper reconciles the two points of view.
\end{abstract}

\maketitle

A major controversy has raged in  Particle Physics recently as to whether the angular momentum (AM) of a photon, and \textit{\`{a} fortiori} a gluon, can be split into  physically meaningful, i.e. measurable, spin and orbital parts. The combatants in this controversy\footnote{For access to the controversy literature see the reviews by Leader and Lorc\'{e} \cite{Leader:2013jra} and Wakamatsu \cite{Wakamatsu:2014zza}. Note, though, that I shall criticize some of the statements in \cite{Leader:2013jra}} seem, largely, to be unaware of the fact that Laser Physicists have been measuring the spin and orbital angular momentum of laser beams for decades!\footnote{For access to the laser literature see the reviews of Bliokh and Nori \cite{Bliokh:2015doa} and Franke-Arnold, Allen and Padgett \cite{LPOR:LPOR200810007}}. My aim is to reconcile these apparently conflicting points of view. Throughout  this paper,  unless explicitly stated, I will be discussing only free fields.\nl

I shall first consider QED, where $\bm{E}, \, \bm{B} $ and $\bm{A}$ are field operators, and as is customary, employ rationalized Gaussian units. It is usually stated that the momentum density in the electromagnetic field (known, in QED, as the \emph{Belinfante} version) is proportional to the Poynting vector, i.e.
\beq \label{Poynting} \bm{P}_{\textrm{bel}}= \int\ud^3x \bm{p}_{\,\text{bel}}(x) \qquad \bm{p}_{\,\text{bel}}(x) = \uvec E\times \uvec B \eeq
and it is therefore eminently reasonable that the AM  should be given by
\beq
\uvec J_\text{bel}= \int\ud^3x \bm{j}_{\,\text{bel}}(x).
 \eeq
 where the Belinfante AM density is
\beq \label{bel}
\uvec j_{ \,\text{bel}}= \uvec r \times (\uvec E \times \uvec B).
 \eeq
 Although this expression has the structure of an orbital AM, \emph{i.e.} $\uvec r\times\uvec p$, it is, in fact, the \emph{total} photon angular momentum density.
 On the other hand, application of Noether's theorem to the rotationally invariant Lagrangian yields the \emph{Canonical} version which has a spin plus orbital part
  \beq
\uvec J_{\text{can}} = \int \ud^3x \,  j_{\,\text{can}} = \int \ud^3x \, [ l_{\,\text{can}} + \uvec s_{\,\text{can}} ] \eeq
where the canonical densities are
 \beq
\uvec s_{\,\text{can}} =  \uvec E \times \uvec A \quad \textrm{and} \quad \uvec l_{\,\text{can}}= E^i (\uvec x \times \uvec \nabla) A^i
\eeq
but, clearly, each term is gauge non-invariant. \nl
Textbooks have long stressed a ``theorem" that such a split cannot be made gauge invariant. Hence the controversial reaction when Chen, Lu, Sun, Wang and Goldman \cite{Chen:2008ag} claimed that such a split \emph{can} be made. They introduce fields $\uvec A_\pure$ and $\uvec A_\phys$, with
\be\label{decomposition}
\uvec A=\uvec A_\pure+\uvec A_\phys  \eeq
where
\beq \label{perppar}  \uvec\nabla\times\uvec A_\pure=\uvec 0, \quad \textrm{and} \quad \uvec\nabla\cdot\uvec A_\phys=0 \eeq
which  are, of course, exactly the same fields as in the Helmholz decomposition into longitudinal and transverse components\footnote{Indeed the only reason for the new nomenclature was Chen et al's  intention to extend these ideas to QCD.}
\beq \uvec A_\pure \equiv \bm{A}_\parallel \qquad \qquad \uvec A_\phys \equiv \bm{A}_\bot. \eeq
Chen et al then obtain
\beq \label{CHEN}
\uvec J_{\text{chen}}=\underbrace{\int\ud^3x\,\uvec E\times\uvec A_\bot}_{\uvec S_\text{chen}}+\underbrace{\int\ud^3x\,E^i(\uvec x\times\uvec\nabla) A^i_\bot}_{\uvec L_\text{chen}} \eeq
and since $ \uvec A_\bot $ and $\bm{E} $ are unaffected by gauge transformations, they appear to achieved the impossible. The explanation is that the ``theorem" referred to above applies to \emph{local} fields, whereas $ \uvec A_\bot $ is, in general, \emph{non-local}. In fact
\beq \label{nonlocal}
\uvec A_\bot(x) = \uvec A(x)     -\frac{1}{4\pi} \uvec\nabla\int \ud^3x' \,\frac{\uvec\nabla'\cdot \uvec A(\uvec x')}{| \uvec x - \uvec x'|}. \eeq
In all three versions of AM just mentioned, the integrands differ by terms of the general form $\bm{\nabla}\cdot \bm{f}$, where $\bm{f}$ is some function of the fields, so that the integrated versions differ by surface terms at infinity, and thus agree with each other if the fields vanish at infinity. For classical fields, to state that a field vanishes at infinity, is physically meaningful, but what does it mean to say an operator vanishes at infinity? The first serious analysis of this question seems to be that of  Lowdon \cite{Lowdon:2014dda}, utilising axiomatic field theory. I shall comment later on his conclusions.\nl
Now the key question is: what is the physical relevance of the various $\bm{S}$  operators? Can they be considered as genuine spin operators for the electromagnetic field? A genuine spin operator should satisfy  the following commutation relations (for an interacting theory these should only hold as ETCs i.e. as Equal Time Commutators)
\beq \label{Scom} [\, S^i \, , \, S^j \,]= i\hbar \epsilon^{ijk} S^k  . \eeq
But to check these conditions, manifestly, one must know the fundamental commutation relations between the fields and their conjugate momenta i.e. the quantization conditions imposed when quantizing the original classical theory, yet to the best of my knowledge, with only one exception \cite{Leader:2011za}, none of the papers in the controversy  actually state what fundamental commutation relations they are assuming. Thus the expressions alone for the operators $\bm{S}$ are insufficient. \nl
Failure to emphasize the importance of the commutation relations in a gauge theory can lead to misleading conclusions. It must be remembered that the quantization of a gauge theory proceeds in three steps:\nl
(1) One starts with a gauge-invariant \emph{classical} Lagrangian. \nl
(2) One chooses a gauge. \nl
(3) One imposes quantization conditions which are compatible with the gauge choice. \nl
I shall comment on just two cases. In covariant quantization (cq)  \cite{Lautrup, Nakanishi:66,Nakanishi:72}, for example in the Fermi gauge, one takes
\beq \label{CovQ} [\, \dot{A}^i(\bm{x},t)\, , \, A^j(\bm{y},t) \,]= -i\delta^{ij}\delta(\bm{x} - \bm{y}),  \eeq
and then the Hilbert space of photon states has  an indefinite metric. \nl
Quantizing in the Coulomb gauge one uses transverse quantization (tq) (see e.g. \cite{Bjorken:1965zz})
\beqy \label{CoulQ} &&[\, \dot{A}^i(\bm{x},t)\, , \, A^j(\bm{y},t) \,]\, =\, -i\delta^\bot_{ij}(\bm{x} - \bm{y})  \\
 &&\label{CoulQ'} \equiv \, -i\int \frac{d^3k}{(2\pi)^3}\left(\delta_{ij} - \frac{k_i k_j}{\bm{k}^2}\right)e^{i\bm{k}\cdot (\bm{x}- \bm{y})} \eeqy
and the Hilbert space is positive-semidefinite.\nl
There is an important physical consequence of this difference in quantization procedures. Gauge transformations on field \emph{operators}  almost universally utilize \emph{classical} functions  i.e.
\beq \label{GT}  \bm{A}(x) \rightarrow \bm{A}(x) + \bm{\nabla} \alpha (x) \eeq
where $\alpha (x)$ is a ``c-number" function. Clearly this transformation cannot alter the commutators. Or, put  another way, gauge transformations are canonical transformations and therefore are generated by unitary operators, which do not alter commutation relations.
This means that one cannot go from say Canonically quantized QED to Coulomb gauge quantized QED via a gauge transformation. This point was emphasized by Lautrup \cite{Lautrup}, who explains that although the theories are physically identical at the classical level, it is necessary to demonstrate that the physical predictions, meaning scattering amplitudes and cross-sections, are the same in the different  quantum versions. This is also stressed by Cohen-Tannoudji, Dupont-Roc and Grynberg \cite*{[][ {; \, English translation: Wiley Professional Paperback Series, 1989}]CohenTannoudji:1987bi} on the basis that also the Hilbert spaces of the different quantum versions are incompatible.\nl
It is not difficult to show that the canonical $\bm{S}_{\textrm{can}}$ with covariant quantization i.e. $\bm{S}^{\textrm{cq}}_{\textrm{can}}$ satisfies   Eq.~(\ref{Scom}) and so is a genuine spin operator. However it is not gauge invariant. I shall comment on this presently.\nl
For the Chen et al case, since we are dealing with free fields, the parallel component of the electric field is zero i.e. $ \bm{E}_\| =0 $ so that $\bm{J}_{\textrm{chen}}$ becomes
\beq \label{chenvEN}\uvec J_{\text{chen}}= \int\ud^3x\,\uvec E_\bot\times\uvec A_\bot + \int\ud^3x\,E_\bot^i(\uvec x\times\uvec\nabla) A^i_\bot. \eeq
But this is exactly the expression for $\bm{J}$, studied, with transverse quantization, in great detail by van Enk and Nienhuis (vE-N) in their classic paper \cite{vanEnk1994}, which is often the basis for statements about spin and orbital angular momentum in Laser Physics i.e one has
\beq \label{vENS} \bm{S}^{\textrm{tq}}_{\textrm{chen}}\equiv\bm{S}^{\textrm{tq}}_{\textrm{vE-N}} = \int\ud^3x\,\uvec E_\bot\times\uvec A_\bot \eeq
and
\beq \label{vENL} \bm{L}^{\textrm{tq}}_{\textrm{chen}}\equiv\bm{L}^{\textrm{tq}}_{\textrm{vE-N}}=\int\ud^3x\,E_\bot^i(\uvec x\times\uvec\nabla) A^i_\bot. \eeq

Now it is clear that $\bm{S}^{\textrm{tq}}_{\textrm{vE-N}}$, and $\bm{L}^{\textrm{tq}}_{\textrm{vE-N}}$, which are gauge invariant, are exactly the same as the gauge-variant canonical versions evaluated \emph{in the Coulomb gauge}. For this reason, following \cite{Leader:2013jra}, we shall henceforth refer to the Chen et al = van Enk-Nienhaus operators  as the Gauge Invariant Canonical (gic) operators. Thus
\beq
\uvec J_{\,\text{gic}}= \uvec L_{\,\text{gic}} + \uvec S_{\,\text{gic}} =   \int\ud^3x\,[ \bm{ l}_{\,\text{gic}} + \bm{ s}_{\,\text{gic}} ] \eeq
where the densities are
\beq  \label{gicdens} \uvec s_{\,\text{gic}} =\uvec E_\bot\times\uvec A_\bot \quad \textrm{and} \quad \uvec l_{\,\text{gic}}= E_\bot^i(\uvec x\times\uvec\nabla) A^i_\bot. \eeq

 Now van Enk and Nienhuis show that the commutation relations for $\bm{S}^{\textrm{tq}}_{\textrm{gic}}$ are very peculiar and  not at all like those in Eq.~(\ref{Scom}).\footnote{Also $\bm{L}^{\textrm{tq}}_{\textrm{gic}}$ has peculiar commutation relations, but as expected, $\bm{J}^{\textrm{tq}}_{\textrm{gic}}$ behaves as a perfectly normal total angular momentum.}  They demonstrate that
\beq \label{vENcom} [\, S^{\textrm{tq},\, i }_{\textrm{gic}}\, , \, S^{\textrm{tq},\, j}_{\textrm{gic}}\,] =0 \,! \eeq
and stress that the components of $\bm{S}^{\textrm{tq}}_{\textrm{gic}}$ \emph{cannot} therefore be considered as the components of a genuine spin vector in general. Moreover,  they are careful to refer to this operator as the `spin' \emph{in inverted commas} (and similarly  $\bm{L}^{\textrm{tq}}_{\textrm{gic}}$, is referred to as the `orbital angular momentum'), but it seems that later papers on Laser Physics have not bothered to respect this convention. \nl
Despite all  these peculiarities it is claimed, correctly,  that the spin and angular momentum of certain types of laser beam can and are regularly measured.\footnote{Similar comments apply also to gluons.} So the key question is how is this to be reconciled with the above, where, on the one hand, we have $\bm{S}^{\textrm{cq}}_{\textrm{can}}$ which looks like a genuine spin operator, but which is not gauge invariant and, on the other, $\bm{S}^{\textrm{tq}}_{\textrm{vE-N}}$, which in no way resembles a spin operator, but which is at least gauge invariant.\nl
It was shown in \cite{Leader:2011za} that  $ (\bm{S}^{\textrm{cq}}_{\textrm{can}}\bm{\cdot} \bm{P}/|\bm{P}|)$, where $\bm{P}$ is the momentum operator,  measures  helicity and that its matrix elements between arbitrary physical photon states are gauge invariant. A key step in this proof was to
 consider the action of $(\bm{S}^{\textrm{cq}}_{\textrm{can}}\bm{\cdot} \bm{P}/|\bm{P}|)$ on the physical photon state $ |\, \bm{k}, \, j \, \ra  $ with transverse polarization. Provided the operators are normal ordered one has
\beq \label{actionS} (\bm{S}^{\textrm{cq}}_{\textrm{can}}\bm{\cdot} \bm{P}/|\bm{P}|)\,|\, \bm{k}, \, j \, \ra =
 \hat{k}^i\,[\, S^{\textrm{cq}, \, i}_{\textrm{can}} \,, \, a^\dag (\bm{k}, j)\,]\,|\textrm{vac} \ra \eeq
 and the commutator is then evaluated using the covariant quantization conditions. Acting on a state of helicty $\lambda$ one eventually finds that $(\bm{S}^{\textrm{cq}}_{\textrm{can}}\bm{\cdot} \bm{P}/|\bm{P}|)$ measures helicity:
 \beq \label{Helres} (\bm{S}^{\textrm{cq}}_{\textrm{can}}\bm{\cdot}  \bm{P}/|\bm{P}|)\,|\bm{k},\lambda \ra = \lambda\hbar\,|\bm{k},\lambda \ra. \eeq
  For the case of the helicity based on  $\bm{S}^{\textrm{tq}}_{\textrm{gic}}$ the analogous commutator has to be evaluated using the transverse commutation conditions Eq.~(\ref{CoulQ}), but it turns out that the terms $k_ik_j $ don't contribute, so that also $\bm{S}^{\textrm{tq}}_{\textrm{gic}}\bm{\cdot} \bm{P}/|\bm{P}|$ measures helicity i.e.
\beq  \label{vEhel} (\bm{S}^{\textrm{tq}}_{\textrm{gic}}\bm{\cdot} \bm{P}/|\bm{P}|)\,|\,\bm{k}, \, \lambda  \,\ra = \lambda \hbar\,|\,\bm{k}, \, \lambda \,\ra. \eeq
 In summary, only the \emph{helicity}, based either on $ \bm{S}_{can}$ or on $\bm{S}_{gic}$, is physically meaningful as a  measure of angular momentum. But, interestingly, as van-Enk and Nienhaus \cite{vanEnk1994} show, the other components of $ \bm{s}_{gic}$, though not angular momenta, are nevertheless measurable quantities. We shall see this concretely in the classical discussion which follows, where, it should be borne in mind that, unlike the QED situation, it is straightforward to compare  expressions in different gauges.\nl

I turn now to the key question which has remained unresolved in the particle physics discussions, namely, which of the AM densities $\bm{j}_{bel}$,  $\bm{j}_{can}$  or
$\bm{j}_{gic}$ is relevant physically. Contrary to the opinion expressed in  \cite{Leader:2013jra}, where it is argued that it is simply a matter of taste, and to \cite{Lowdon:2014dda}, which favours the Belinfante version,  I shall argue that the laser experiments clearly indicate that it is $\bm{j}_{gic}$ which plays a direct role in the interaction of  classical EM waves with matter and that the Belinfante expression is definitely unacceptable. The criticism that a density should  not depend on a non-local field $\bm{A}_\bot $ does not apply to the situation of most interest, namely when dealing with  monochromatic free fields with time dependence $e^{-i\omega t}$, since then    $\bm{E} = \bm{E}_\bot= - \dot{\bm{A}}_\bot $ so that
   \beq \label{Aperp} \bm{A}_\bot = -\frac{i}{\omega}\, \bm{E}  \eeq
   is a local field. \nl
Discussing the  classical electrodynamics of laser fields, I shall follow custom and switch to SI units. \emph{The only effect on all the previous formulae for momentum and AM densities is to multiply them by a factor $\epsilon_0$.}\nl
The real, monochromatic physical EM fields $( \bm{\mc{E}}, \bm{\mc{B}}) $ are, as usual, expressed in terms of  complex fields $( \bm{E},\bm{B})$
\beq \bm{\mc{E}} = Re( \bm{E})  \qquad \bm{E}(\bm{r},t) = \bm{E}_0(\bm{r})\,e^{-i\omega t} \eeq
\beq \bm{\mc{B}} = Re( \bm{B})  \qquad \bm{B}(\bm{r},t) = \bm{B}_0(\bm{r})\,e^{-i\omega t}. \eeq
The force on, and the torque (about the centre of mass of a small neutral object), in dipole approximation, are given by
\beq \label{force} \bm{F} = (\bm{\mc{P}}.\bm{\nabla})\,\bm{\mc{E}} + \dot{\bm{\mc{P}}}\bm{\times}\bm{\mc{B}} \qquad \quad \bm{\mc{\tau}}= \bm{\mc{P}}\bm{\times}\bm{\mc{E}} \eeq
where the induced electric dipole moment is given by
\beq \label{dipmom} \bm{\mc{P}}=  Re[\alpha \bm{E}(\bm{r},t)] \eeq
and the complex polarizability is
\beq \alpha = \alpha_R  +  i \, \alpha_I .\eeq
First consider the force acting on the neutral dipole.
 In \cite{PhysRevA.88.033831} it is shown that the total force splits into two terms
  \beq \label{splitf} \bm{F} = \bm{F}_{\textrm{reactive}}  + \bm{F}_{\textrm{dissipative}}  \eeq
  where, for the cycle average,  which I indicate by $< \, > $,
  \beq \label{fdiss} \langle \bm{F}_{\textrm{dissipative}}  \rangle = \frac{ \alpha_I}{2}Im[E^{*i} \bm{\nabla} E^i] \eeq
   and for a classical electric dipole with momentum $\bm{P}_{\textrm{dipole}}$ it is    $\bm{F}_{\textrm{dissapative}}$ that controls its rate of change of momentum (see Chapter V of \cite{CohenTannoudji:2004})
  \beq \label{momchng} \Big\langle   \frac{d\bm{P}_{\textrm{dipole}}}{dt}  \Big\rangle = \langle \bm{F}_{\textrm{dissapative}} \rangle . \eeq
  Naturally, for the linear momentum, as for the AM,  besides the Belinfante version Eq.~(\ref{Poynting}), there exist also the gauge-variant canonical and gauge-invariant gic versions
 \beq \label{Pcan}        \bm{P}_{\textrm{can}}= \epsilon_0\int\ud^3x \,  \mc{E}^i \bm{\nabla} \mc{A}^i \eeq
  and
  \beq \bm{P}_{\textrm{gic}}= \int\ud^3x \,\bm{p}_{\textrm{gic}} \quad \textrm{with} \quad  \bm{p}_{\textrm{gic}} = \epsilon_0 \, \mc{E}^i \bm{\nabla} \mc{A}^i_\bot \eeq
  and as in the AM case the three space-integrated versions are equal if the fields vanish at infinity.\nl
  Evaluating the cycle average, using Eq.~(\ref{Aperp}), it turns out that
  \beq \label{Fdisscyc} \langle \bm{F}_{\textrm{dissipative}}  \rangle =  \frac{\alpha_I \omega}{\epsilon_0} \, \langle \bm{p}_{\textrm{gic}}  \rangle \eeq
  so that  it is the gauge-invariant canonical version that is physically relevant, and it is, of course, equal to the canonical version evaluated in the Coulomb gauge.\nl
Next consider the torque about the centre of mass of the dipole. One finds that
\beq \label{P} \bm{\mc{P}}=  \alpha_R \,\bm{\mc{E}} - \frac{\alpha_I}{\omega}\, \dot{\bm{\mc{E}}} \eeq
so that
\beq \label{tau} \bm{\mc{\tau}}=\frac{\alpha_I}{\omega}\, \bm{\mc{E}} \bm{\times} \dot{\bm{\mc{E}}}. \eeq
For the cycle average, one finds
\beq \label{tave} \langle \bm{\mc{\tau}} \rangle = \alpha_I [ Re\bm{E}_{0} \bm{\times} Im\bm{E}_{0}]. \eeq
Now consider the cycle average of  $\uvec s_{\,\text{gic}}$ given in Eq.~(\ref{gicdens})
\beqy \label{scycave} \langle \bm{s}_{\textrm{gic}}\rangle &= &\frac{1}{2\omega}\, \epsilon_0\, Im[\bm{E}^* \bm{\times} \bm{E} ] \nn \\
&&= \frac{1}{\omega} \, \epsilon_0\, [Re\bm{E}_0 \bm{\times} Im\bm{E}_0] \eeqy
so that from Eq.~(\ref{tave}) follows the fundamental result
\beq \label{taufinal} \langle \bm{\mc{\tau}} \rangle =\frac{\alpha_I \omega}{ \epsilon_0} \langle \bm{s}_{\textrm{gic}}\rangle . \eeq
The physical torque is thus given by a gauge-invariant expression, as it ought to be, which coincides with the canonical version evaluated in the Coulomb gauge, in accordance with the entire discussion in \cite{Bliokh:2015doa}. At first sight it may seem odd that only the spin vector enters Eq.~(\ref{taufinal}), but it should be remembered that $\bm{\tau}$ is the torque about the centre of mass of the dipole, whereas $\bm{L}$ is the orbital AM about the origin of the axis system.\nl

Consider now the application of these results to lasers. In  the foundation paper on laser angular momentum by Allen, Beijersbergen, Spreeuw and Woerdman \cite{PhysRevA.45.8185}   the AM is associated with the gauge invariant Belinfante version in Eq.~(\ref{bel}). It is therefore important to review some of the  properties of the AM density $\bm{j}_{\textrm{bel}}$ and  of the Belinfante linear momentum, whose density is proportional to the Poynting vector. Firstly, for a plane wave propagating in the $Z$-direction the helicity is the same as the $z$-component of the angular momentum, and, as shown in Section 2.6.4 of \cite{Leader:2013jra}, for a left-circularly polarized i.e. positive helicity beam, $ \bm{j}_{\textrm{bel}, \, z}=0$, whereas, \emph{per photon}
  \beq   \bm{j}_{\textrm{can}, \,z} =\bm{s}_{\textrm{can}, \, z} = \bm{j}_{\textrm{gic}, \, z} = \bm{s}_{\textrm{gic}, \, z} =\hbar \eeq
  as intuitively expected. Moreover, this result is much more general: $\bm{j}_{\textrm{bel}}$ obviously has zero component in the direction of the Belinfante field momentum density:
  \beq \label{jpzero} \bm{j}_{\textrm{bel}}\cdot \bm{p}_{\textbf{bel}} =\epsilon_0^2 \,\left[\uvec r \times (\bm{\mc{E}} \bm{\times} \bm{\mc{ B}}) \right] \cdot (\bm{\mc{E}} \bm{\times} \bm{\mc{ B}}) =0 .\eeq
  Thus the Belinfante AM fails, whereas the gauge invariant canonical version succeeds, in correctly generating the helicity. Secondly, and this seems most surprising in light of the initial comments on the controversy given above, it will be seen presently that for a superposition of polarized plane waves, $\bm{j}_{\textrm{bel}}$ splits into two terms apparently corresponding to orbital and spin angular momentum   \cite{PhysRevA.45.8185}. \nl

In their analysis Allen et al utilize the paraxial approximation, which corresponds to keeping the first two terms in an expansion in terms of a parameter equal to the beam waist divided by the diffraction length  \cite{PhysRevA.11.1365}, and apply it to a Laguerre-Gaussian laser mode, but their treatment is actually more general and applies to any monochromatic, axially-symmetric vortex beam of finite cross-section. In such a beam propagating in the $Z$-direction the complex electric field, in paraxial approximation and in the notation often used in laser papers, has the form\footnote{Often the $z$-component is put equal to zero, but that gives zero for the Belinfante angular momentum, whereas the laser papers have the non-zero value obtained below.}
  \beq \label{Epara} \bm{E}= i\omega \,\Big(u(\bm{r}), v(\bm{r}), \frac{-i}{k}\big(\frac{\partial u}{\partial x} + \frac{\partial v}{\partial y}\big)\Big)e^{i(kz-\omega t)} \eeq
  where
  \beq \Big|\frac{\partial u}{\partial z}\Big|  \ll k|u|  \qquad \Big|\frac{\partial v}{\partial z}\Big|  \ll k|v| \eeq
   and all second derivatives and products of first derivatives are ignored. As in \cite{Bliokh:2015doa} I shall indicate relations that are  valid in paraxial approximation by "$\simeq $", so for example $\omega \simeq kc $. \nl
   For the case of circularly polarization
   \beq v= i \sigma_z u \eeq
   where $\sigma_z= \pm 1 $ for left/right circular polarization, and in cylindrical coordinates $(\rho, \, \phi, \, z)$
   \beq \label{u} u(\rho,\phi,z)= f(\rho, z) e^{il\phi}. \eeq
   For the cylindrical components of  the cycle averaged Belinfante momentum density  one finds
     \beq \label{pbelcyc} \langle p_{\textrm{bel}} \rangle _\rho \simeq -\epsilon_0 \omega\, Im\big(u\frac{\partial u^*}{\partial \rho}\big)  \qquad
   \langle p_{\textrm{bel}} \rangle _z \simeq \epsilon_0 k \omega | u |^2 \eeq
   \beq \label{pphi} \langle p_{\textrm{bel}} \rangle _\phi \simeq \epsilon_0 \omega\left[ \frac{l}{\rho}|u|^2 -\frac{\sigma_z}{2}\frac{\partial |u|^2 }{\partial\rho}\right]. \eeq
  Most interesting is the $z$-component of the Belinfante AM density\footnote{Note that this does not contradict Eq.~(\ref{jpzero}) since $\bm{j}_{\textrm{bel}}$ does not point along the $Z$-direction.}
   \beqy \label{jbelz} \langle j_{\textrm{bel}} \rangle _z &=& [ \bm{r} \bm{\times} \langle p_{\textrm{bel}} \rangle]_z = \rho \langle p_{\textrm{bel}} \rangle _\phi \nn \\
   &\simeq &  \epsilon \omega\left[ l|u|^2 -\frac{\sigma_z}{2}\rho\frac{\partial |u|^2 }{\partial\rho}\right],
         \eeqy
         implying the unintuitive result that \emph{per photon}
 \beq \label{jbelzphot} \langle j_{\textrm{bel}} \rangle ^{\textrm{photon}}_z \simeq l\hbar -\frac{\sigma_z\hbar}{2|u|^2}\rho \frac{\partial |u|^2}{\partial \rho}. \eeq
 On the contrary for the gauge invariant canonical version one finds
 \beq \label{jgicpara}  \langle l_{\textrm{gic}} \rangle \simeq \epsilon_0 \omega l|u|^2 \qquad \langle s_{\textrm{gic}} \rangle\simeq  \epsilon_0 \omega \sigma_z|u|^2 \eeq
 implying the beautiful result \emph{per photon}
 \beq \label{jgicphot} \langle l_{\textrm{gic}} \rangle ^{\textrm{photon}}_z \simeq l\hbar \qquad
 \langle s_{\textrm{gic}} \rangle ^{\textrm{photon}}_z \simeq \sigma_z\hbar.  \eeq
 Not surprisingly, if one integrates Eq.~(\ref{jbelz}) over the beam cross-section, one obtains for the average $ \langle j_{\textrm{bel}} \rangle_z $ \emph{per photon}
 \beq  \langle j_{\textrm{bel}} \rangle ^{\textrm{photon}}_z \Big|_{ave}\simeq l\hbar  + \sigma_z\hbar. \eeq
 However, crucially, for small enough dipoles the angular  momentum absorbed depends on the \emph{local} AM density, which, comparing Eqs.~(\ref{jbelzphot}, \ref{jgicphot}) is quite different for the Belinfante and gic cases, even differing in sign between the beam axis and the beam periphery. The first semi-quantitative test of the above was made by Garc\'{e}s-Ch\'{a}vez, Mc Gloin, Padgett, Dulz, Schmitzer and Dholakia \cite{PhysRevLett.91.093602} who succeeded in studying the motion of  a tiny  particle trapped at various radial distances $\rho$  from the axis of a so-called Bessel beam. The transfer of orbital AM causes the particle to circle about the beam axis with a rotation rate $\Omega_{\textrm{orbit}}$ whereas the transfer of spin AM causes the particle to spin about its centre of mass with rotation rate $\Omega_{\textrm{spin}}$. Given that, for a Bessel beam, $|u|^2 \propto 1/\rho $  one finds for the Belinfante case that
 \beq \label{belomegas} \Omega_{\textrm{orbit}}\propto 1/\rho^3    \qquad    \textrm{and} \qquad \Omega_{\textrm{spin}}\propto 1/\rho, \eeq
 which is precisely the behaviour found experimentally, apparently showing  the the Belinfante expressions are the correct physical ones. However, exactly the same functional dependence on $\rho$ follows from the gic expressions. In fact this equivalence is not restricted to Bessel beams. It holds as long as $|u|^2$ follows a simple power law behaviour $|u|^2 \propto \rho^{-\beta }$. Since the absolute rotation rates depend upon detailed parameters which, according to the authors,  were  beyond experimental control, it would be incorrect to interpret these results as evidence in favour of the Belinfante expressions. Moreover, in an unpublished paper \cite{XSChen},  Chen and Chen have argued that the dependence on $l$ and $\sigma_z$, of the shift of the diffraction fringes, found by Ghai, Senthilkumaran and Sirohi \cite{Ghai:2009}   in the single slit diffraction of optical beams with a phase singularity,  implies that the correct expression for the optical  angular momentum density is the gic one. And, further, as summarized in the recent review \cite{Bliokh:2015doa} it is the canonical AM in the Coulomb gauge i.e the gic AM that agrees with a wide range of experiments.\nl
  For the linear momentum, on the other hand, it seems more difficult to choose experimentally between the Belinfante and gic versions, but I shall give an argument in favour of the gic version for photons. For the cycle averages one finds
\beq \label{belgicmom} \langle \bm{p}_{\textrm{bel}} \rangle = \langle \bm{p}_{\textrm{gic}} \rangle + \frac{\epsilon_0\omega}{2}Im[(\bm{E}\cdot \bm{\nabla})\bm{E}^*]. \eeq
and in the paraxial case under discussion this becomes
\beq \label{parbg} \langle \bm{p}_{\textrm{bel}} \rangle_{\textrm{paraxial}} = \langle \bm{p}_{\textrm{gic}} \rangle_{\textrm{paraxial}}- \frac{\epsilon_0\omega\sigma_z}{2}\frac{\partial|u|^2}{\partial\rho} \hat{\phi}. \eeq
Following \cite{CohenTannoudji:2004}, assuming that the change of momentum of the dipole is due to the momentum of the photons absorbed from the beam, I shall take the number of  photons totally absorbed by the dipole per second to be given by   $1/\hbar\omega$ times the rate of increase of the dipole's internal energy. For a paraxial beam  I then find that Eqs.~(\ref{Fdisscyc}) and (\ref{momchng}) are satisfied only if the average photon momentum is taken as
  \beq \label{photmomav} \langle \bm{p} \rangle ^{\textrm{photon}} \Big|_{ave} \simeq \frac{1}{N}\langle \bm{p}_{\textrm{gic}} \rangle \eeq
  where $N$ is the number of photons per unit volume. A similar argument supports the gic version for the AM. Namely,  assuming that the change in internal angular momentum of the dipole arises from photon absorption I find that Eq.~(\ref{taufinal}) is satisfied only if
  \beq \label{photspinav} \langle \bm{s} \rangle ^{\textrm{photon}} \Big|_{ave} \simeq \frac{1}{N}\langle \bm{s}_{\textrm{gic}} \rangle. \eeq

  In summary, the angular  momentum controversy, which has bedevilled particle physicists for some time, is resolved by a host of laser physics experiments which indicate that the Gauge Invariant Canonical linear momentum and angular momentum densities are the physically relevant ones, and that this is not simply
   \newpage
   a question of taste.\footnote{ This is in contradiction with Lowdon \cite{Lowdon:2014dda} who favours the Belinfante expressions. Starting with Belinfante   he  finds no reason to expect  that the operator surface terms  vanish, as would be necessary in order to make the Canonical AM equal to the Belinfante AM. However, if he had \emph{started} with  the Canonical AM he would have reached the opposte conclusion.} Moreover,  although there does not exist a genuine spin vector for photons, the van Enk-Nienhuisen=Chen et al= gic `spin vector'   plays a central role in Laser Physics. All of its components can, in principle, be measured, but only one component, strictly speaking the helicity, is a genuine AM. For   a paraxial beam propagating in the $Z$-direction one can show that the $Z$-component of the gic spin vector coincides with the gic helicity i.e $ \langle S_{\textrm{gic}} \rangle_z \simeq \langle \textrm{gic helicity} \rangle $, so  this component is effectively a genuine AM. And finally, recognizing that the fundamental expressions are the gic ones, allows one to avoid the somewhat disturbing claim that what is physically measured  corresponds to a gauge-variant quantity evaluated in a particular gauge, i.e. the Coulomb one. \nl

  I am grateful to K. Bliokh for making me aware of the vast literature on laser angular momentum, to  X-S. Chen, C. Lorc\'{e} and G Nienhuis for helpful comments, and to J.Qiu and R Venugopalan for hospitality at Brookhaven National Laboratory. I thank  the Leverhulme Trust for an Emeritus Fellowship.

\bibliography{Elliot_General}
\end{document}